# Improved Drude-Lorentz dielectric function for gold nanospheres

Anthony Centeno
School of Engineering, University of Glasgow
Glasgow, United Kingdom. G12 8QQ

Email: Anthony.Centeno@glasgow.ac.uk

**Abstract:** There is an ever-increasing body of research in applying the Localized Surface Plasmon Resonance (LSPR) of gold nanospheres for applications in the visual part of the electromagnetic spectrum. An important aspect of this work is the accurate representation of the dielectric function for nanospheres with diameters less than the mean free path of the conduction electrons in the bulk material. This paper describes the development of a modified Drude-Lorentz dielectric functions for bulk gold, and for nanosphere. The models are validated against previously published experimental work and are shown to be valid for spheres of diameters of greater than 10 nm.

## 1. Introduction

The interaction of Gold (Au) with electromagnetic waves is widely researched in the visible and near infra-red (VIS-NIR) spectrum. The main reasons for this are that Au shows VIS-NIR resonant interaction and has good biocompatibility compared to silver [1]. This makes Au especially suitable for medical applications.

The resonant interaction is due to surface plasmons being supported at the metal dielectric interface. At resonance there is a very large electric field enhancement causing intensive absorption and scattering of light [1]. Many applications of surface plasmon resonance have been investigated including: sensing [2], surface-enhanced Raman spectroscopy [3-5], surface-enhanced infrared absorption spectroscopy [6], enhanced nonlinear wave mixing [7] plasmonic antenna [8], optoelectronic devices [9], optical metamaterials [10], solar cells [11], carbon reduction [12], solar hydrogen[13].

For Au metal nanoparticles a localized surface plasmon resonance (LSPR) can be excited by incident visual light. The LSPR is strongly dependent on the size, geometry, dielectric environment, and separation distance between adjacent nanoparticles. The spectral properties of the LSPR of the noble metal nanoparticles, gold and silver, have been extensively researched and the extinction properties calculated using Mie theory [14] . For nanoparticles with dimensions above 20 nm the LSPR extinction peak experiences a red shift with increasing sizes due to retardation effects and increasing contributions from multipolar terms.  For dimensions above 50 nm, radiative damping of the collective electronic excitation (i.e. large scattering component of extinction) broadens the line width of the extinction. In these instances, the measured bulk optical properties of the metal can be used for the design and analysis of the LSPR of noble metal nanoparticles [15-19].

When the nanoparticle diameter is less than 30 nm, the intrinsic size-effects become important and accounts for size dependent modifications of the effective permittivity [15] . Most notably an increase in surface damping by very small particles will cause an increase in the bandwidth of the LSPR response [15].

In this work two areas are considered. The first is the fitting of a Drude-Lorentz (DL) model to measured data for bulk Au, using both global and local optimization. Once a good fit is obtained size dependent modifications are introduced. It is found that using measured data from Olman et al [23] to obtain the DL model for bulk material the measured plasmonic response of 33 nm Au nanospheres could be accurately reproduced using a Mie scattering algorithm [20, 21]. For spheres with diameter less than 30 nm the accuracy was further improved by correcting for the effects of increased surface damping. However, for diameters less than 10 nm it is seen that the DL Lorentz model does not accurately predict the LSPR response.

### 1.1. Material Model for Bulk Au

It is well known that the dielectric function can be expressed in the following form [22]:

$$\varepsilon_r(\omega) = \varepsilon_r^{(f)}(\omega) + \varepsilon_r^{(b)}(\omega) \qquad (1)$$

Equation 1 explicitly separates the free electron effects (intraband) from the bound electron effects (interband). The intraband part can be described by the well known Drude model:

$$\varepsilon_r^{(f)} = 1 - \frac{\Omega_p^2}{\omega^2 - i\omega\Gamma_0} \qquad (2)$$

Where $\Omega_p$ is the bulk plasma frequency, $\omega$ is the angular frequency, $i$, is the complex operator and $\Gamma_0$ is the bulk damping rate.

The interband part of the dielectric function is described by Lorentz oscillators:

$$\varepsilon_r^{(b)}(\omega) = \sum_{j=1}^{n} \frac{f_j \omega_p^2}{\omega_{o,j}^2 - \omega^2 - i\omega\Gamma_j} \qquad (3)$$

Where $j$ is the number of oscillators with $\omega_{o,j}$ frequency, $\Gamma_j$ is the damping term associated with the $j$th oscillator and $\sqrt{f_j}\omega_p$ is the plasma frequency associated with the intraband transition, $f_j$ being the oscillator strength. Equations 3 and 2 can now be inserted into Equation 1 to give:

$$\varepsilon_r(\omega) = 1 - \frac{f_0 \omega_p^2}{\omega^2 - i\omega\Gamma_0} + \sum_{j=1}^{n} \frac{f_j \omega_p^2}{\omega_{o,j}^2 - \omega^2 - i\omega\Gamma_j} \qquad (4)$$

Where $\Omega_p = f_0 \omega_p^2$.

Data for refractive index of Au can be obtained experimentally by ellipsometry, and a number of datasets are available on the refractive index database [23](https://refractiveindex.info). To fit the experimental data to the Drude-Lorentz model given in equation (4) a global optimization technique can be used, for example the simulated annealing method [22, 24] or the differential evolution method [25]. A global optimization technique is preferable to using a least square algorithm, which is very sensitive to the initial guess. In this work Differential Evaluation (DE) and Dual Annealing (DA) global optimization are implemented. This was accomplished by modifying the code originally developed by Juluri [25] for DE to also run a DA algorithm in Python 3.

The Drude-Lorentz model used consisted of one Drude and five Lorentz oscillator terms. Once Drude-Lorentz parameters are obtained these were then used as initial guesses in a Least-Squares (Levenburg-Marquardt) optimization algorithm to obtain the final parameters. Fits were obtained for datasets of bulk Au on the refractive index database. These were validated using Mie scattering calculations of extinction and absorption for small Au spheres.

### 1.2. Material Model for Small Diameter Au Spheres

The classical approach for describing the size dependence of the dielectric function assumes that if the particle size is smaller than the mean free path of the conduction electrons in the bulk, the scattering of the conduction electrons from the particle surface results in a reduced effective mean free path, $L_{eff}$, and an associated increase in bulk damping rate [16]. Equation 2 can now be written as:

$$\varepsilon_r^{(f)} = 1 - \frac{\Omega_p^2}{\omega^2 - i\omega\Gamma_{Leff}} \tag{5}$$

Where:

$$\Gamma_{Leff} = \Gamma_0 + A\frac{v_F}{L_{eff}} \tag{6}$$

*A* being a dimensionless parameter and $v_F$ the Fermi velocity (1.4 nm.fs$^{-1}$ for Au). The parameter *A* accounts for additional surface damping terms [15].

Coronado and Shahtz [16] proposed a geometrical probability measure for calculating $L_{eff}$ of arbitrary shape convex particles, where:

$$L_{eff} = \frac{4V}{S} \tag{7}$$

Where V is the volume and S the surface area of the particle, respectively.

For a sphere equation (7) gives:

$$L_{eff} = \frac{4a}{3} \tag{8}$$

Where a is the radius of the sphere.

Equation 4 can now be modified accordingly to:

$$\varepsilon_r(\omega) = 1 - \frac{f_0\omega_p^2}{\omega^2 - i\omega\Gamma_{eff}} + \sum_{j=1}^{n}\frac{f_j\omega_p^2}{\omega_{o,j}^2 - \omega^2 - i\omega\Gamma_j} \tag{9}$$

### 1.3. Mie scattering calculations of extinction, absorption and scattering for small spheres.

To extract the parameter $\Gamma_{eff}$ experimental data for the extinction ($Q_{ext}$) and absorption efficiency ($Q_{abs}$) of Au spheres is considered. The extinction efficiency can be calculated using Mie theory [21]:

$$Q_{ext} = \frac{2}{x^2}\sum_{n=1}^{\infty}(2n+1)\operatorname{Re}(a_n + b_n) \tag{10}$$

and the absorption efficiency from:

$$Q_{abs} = Q_{ext} - Q_{sca} \tag{11}$$

Where:

$$Q_{sca} = \frac{2}{x^2}\sum_{n=1}^{\infty}(2n+1)(|a_n|^2 + |b_n|^2) \tag{12}$$

x=ka, k being the wavenumber, $k = 2\pi/\lambda$ with $\lambda$ being the wavelength. $a_n$ and $b_n$ are Mie coefficients given by:

$$a_n = \frac{m^2 j_n(mx)[xj_n(x)]' - j_n(x)[mxj_n(mx)]'}{m^2 j_n(mx)[xh_n^{(1)}(x)]' - h_n^{(1)}(x)[mxj_n(mx)]'} \tag{13}$$

$$b_n = \frac{j_n(mx)[xj_n(x)]' - j_n(x)[mxj_n(mx)]'}{j_n(mx)[xh_n^{(1)}(x)]' - h_n^{(1)}(x)[mxj_n(mx)]'} \tag{14}$$

Where m is the refractive index of the sphere relative to the ambient medium and the functions $j_n(x)$ and $h_n^{(1)}(x)$ are Riccati-Bessel functions. In equations (10) the infinite series are truncated when $n_{max} = x + 4x^{1/3} + 2$ [21]. In this investigation PyMieScatt, an open-source Mie computational package, was used to calculate $Q_{ext}$ and $Q_{abs}$ [26].

## 2. Method for Determining Drude-Lorentz Dielectric Function for bulk and small sphere Au.

The Drude-Lorentz (DL) function for bulk Au is obtained by taking experimental data and fitting to the DL dielectric function (Equation 4) using global and least-square optimization. The DL function considered had one Drude term and five Lorentz oscillators. This was chosen since good fits had been reported with this number of Lorentz oscillators in previous literature [22]. In this work Differential Evaluation (DE) and Dual Annealing (DA) global optimization are implemented by modifying code originally developed by Juluri [25]. The software makes extensive use of the Lmfit package (https://github.com/lmfit ). The initial Drude-Lorentz parameters obtained from global optimization were used as initial guesses in a Least-Squares (Levenburg-Marquardt) algorithm to obtain the final DL parameters.

Experimental data from various sources has been made available on the refractive index database [23]. In this work data sets from Johnson and Christy [27] Olman et al [28], McPeak et al [29] and Werner et al [30] were used.

Once a DL model for bulk Au is obtained that provides a good fit to measured data it was modified for small diameter Au spheres. This was accomplished by calculating absorption or extinction as detailed in section 1.3. The dielectric constant used in the Mie calculations was derived using equation (9). Several calculations were carried out, with iterations of the dimensionless parameter *A*, until a good fit between the calculation and experimental data was observed. The experimental data was obtained from references [15, 31] for spheres of diameter between 5 nm and 33 nm.

## 3. Results and Discussion
### 3.1. Drude-Lorentz parameters for bulk Au

The optimization results obtained are shown in Figures 1-4, where the green circles are the experimental data, the red line is the global optimization, and the blue line is the least squares optimization. The best fit to measured data was obtained for the data in reference [28], Olman et al (Figure 1). Since, in this work, the LSPR of small Au spheres are of interest. a good fit over the visual part of the electromagnetic spectrum, 1.65-3.25 eV was considered essential. This is due to it being the region of the electromagnetic spectrum where the Localized Surface Plasmon Resonance of Au nanoparticles are seen.

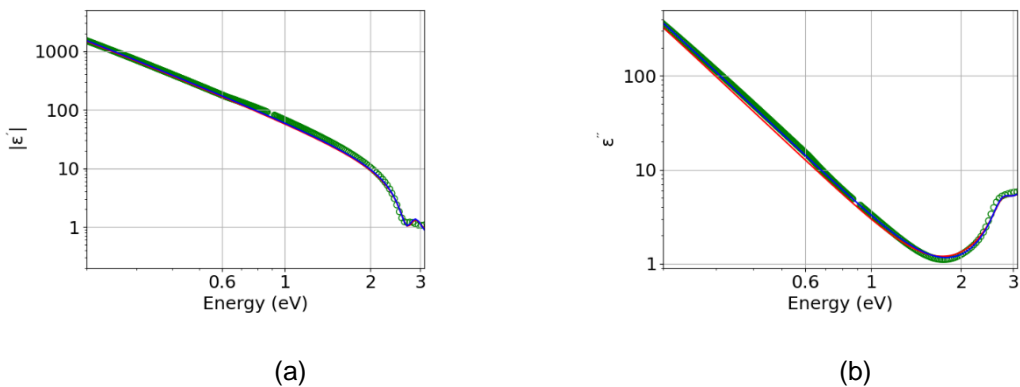

(a)  (b)

*Figure 1. Optimized fit to experimental data from Olman et al [28] for (a) real part and (b) imaginary part of dielectric constant.*

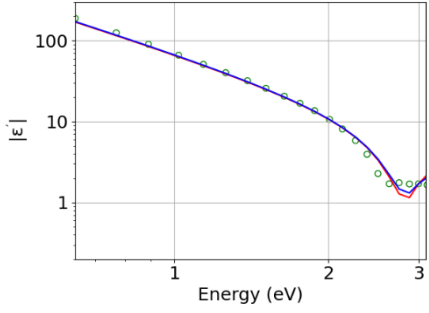
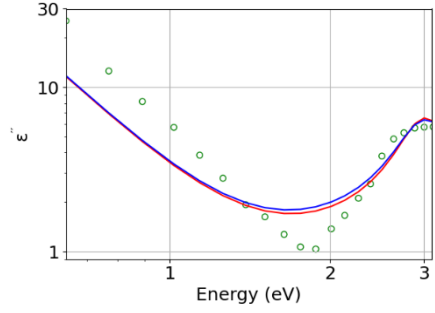

(a)  (b)

Figure 2. Optimized fit to experimental data from Johnson and Christy [27] for (a) real part and (b) imaginary part of dielectric constant.

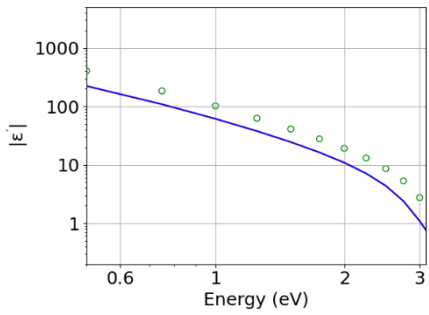
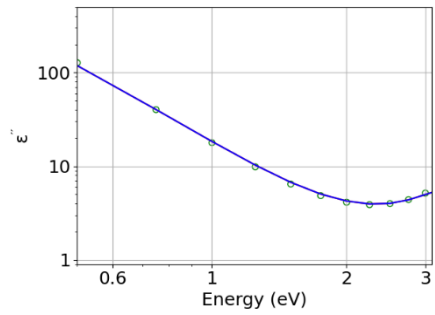

(a)  (b)

Figure 3. Optimized fit to experimental data from Werner et al [31] for (a) real part and (b) imaginary part of dielectric constant.

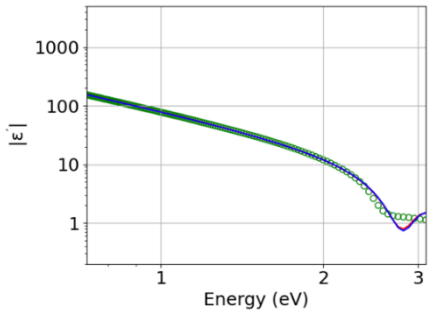
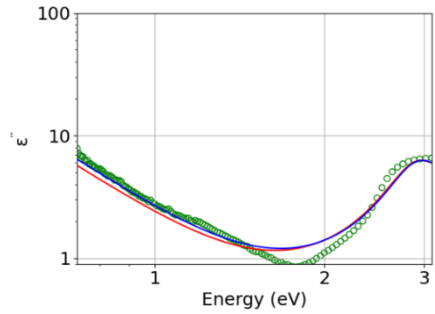

(a)  (b)

Figure 4. Optimized fit to experimental data from McPeak et al [29] for (a) real part and (b) imaginary part of dielectric constant.

The DL fit obtained from the experimental data of reference [28] (Figure 1) was used to develop the modified model for small diameter Au spheres. The parameters for the Drude-Lorentz model for bulk Au are given in Table 1.

| Gold $\omega_p = 8.45$ | | | | | | |
|---|---|---|---|---|---|---|
| $i$ | 0 | 1 | 2 | 3 | 4 | 5 |
| $f$ | 0.98 | 0.1222 | 0.2492 | 2.7845 | 0.1082 | 7.8214 |
| $\omega_o$ | 0 | 3.5684 | 4.2132 | 9.1499 | 2.9144 | 38.9633 |
| $\Gamma$ | 4.4113x10$^{-2}$ | 9.7329x10$^{-1}$ | 1.1139 | 4.637x10$^{-1}$ | 7.0308x10$^{-1}$ | 4.8978x10$^{-1}$ |

Table 1. Drude-Lorentz parameters (in eV) obtained from the optimization fit to data from [28]

### 3.2. Dielectric function for small Au spheres

The DL model was now applied to the problem of calculating the plasmonic response of small diameter Au spheres. This was accomplished by using experimental data in reference [15] for 33 nm, 10 nm and 5 nm and reference [31] for 21.5 nm and 11.3 nm diameter spheres, respectively.

Initially a comparison was made with DL parameters that are used in the popular open source FDTD software MEEP [32]. This DL model was derived by Rakić et al [22] using a thermal annealing optimization algorithm to fit to experimental data from Dold and Mecke [33] and Thèye [34]. Figure 5 shows the results obtained from calculating the absorption for a Au sphere of 33 nm diameter using the Rakić DL parameters, and those derived in this work (shown in Table 1). The experimental results are from reference [15], where the absorption in the graphs is normalized to the highest value of $Q_{abs}$ between 0.5 and 0.6 µm. The normalization was carried out because the experimental results had arbitrary units with a maximum of 1, so this enabled a direct comparison to be made.

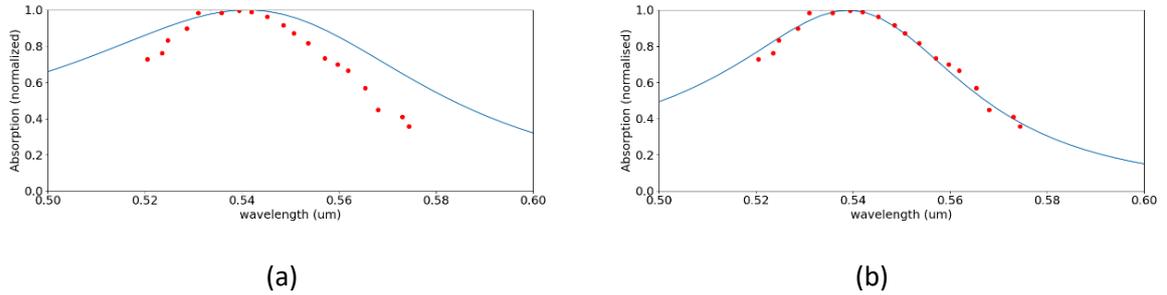

(a)                  (b)

*Figure 5. Comparison of calculated and measured absorption for a 33 nm diameter Au sphere using DL parameters from (a) Rakić et al [22] and (b) Table 1. The dielectric constant of the surrounding medium is 2.2. Experimental data from [15].*

From Figure 5 it can be seen that the DL parameters obtained in this work provide a more accurate calculation of $Q_{abs}$. In both cases the peak of the LSPR is accurate, but the bandwidth of the LSPR peak in Figure 5(a), for the parameters obtained from [28], appears to be significantly broader than the measured data. It should be noted the value of Γ used in the Drude term did not require modification in this instance (i.e. equation (4) is used to find $\varepsilon_r(\omega)$ used in the Mie calculation of absorption in Figure 5(b)).

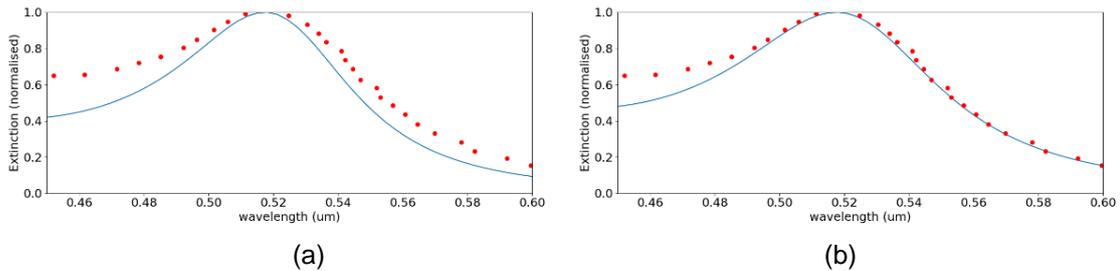

(a)                  (b)

*Figure 6. Comparison of calculated and measured extinction for a 11.3 nm diameter Au sphere using (a) unmodified and (b) modified DL parameters. The dielectric constant of the surrounding medium is 1.7. Experimental data from [31]*

Au spheres of smaller diameter are now considered. Using measurements for normalized extinction efficiency, for 11.3 nm diameter Au spheres obtained from reference [31], comparisons were made using the dielectric function given in equation (4) (Figure 6(a)). The result shows that the bandwidth of the LSPR is narrower than measured, indicating that surface damping is significant for this diameter sphere. Figure 6(b) shows a corrected DL model (Equation 9), where A=0.45 and shows a much-improved fit to the measured data.

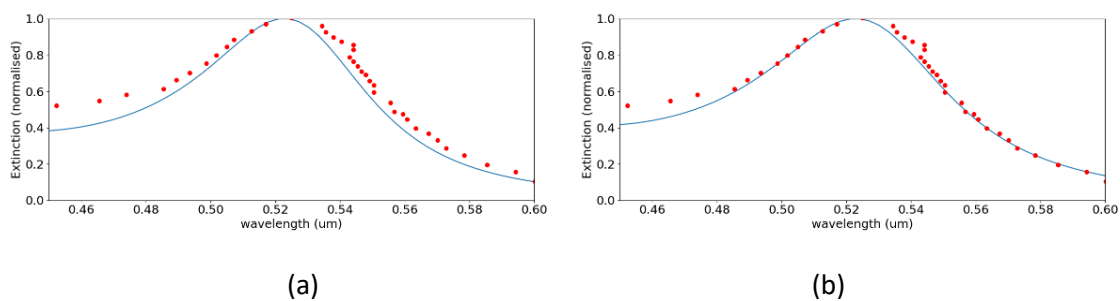

*Figure 7. Comparison of calculated and measured extinction for a 21.5 nm diameter Au sphere using (a) unmodified and (b) modified DL parameters. The dielectric constant of the surrounding medium is 1.7. Experimental data from reference [32]*

Figure 7 shows results for a Au sphere of diameter 21.5 nm. As with the 11.3 nm diameter spheres, the correction to the DL parameters improved the fit to measured data when a value for A of 0.45 is used. However, it is noticeable that the difference between the unmodified and modified case is much less pronounced than the 11.3 nm diameter results in Figure 6. This is to be expected as the effects of surface damping will reduce with increasing diameter.

These results demonstrate the need to account for increased surface damping for spheres of diameters less than 30 nm. However, it might be expected that there is a minimum dimension for which the simple correction to the bulk damping rate would maintain accuracy. Indeed, it was shown by Karimi et al [31] that there is a blue shift in the LSPR of small particles due to quantum size effects. In this work the modified DL dielectric function (Equation 6 with a value of A 0f 0.45) was applied to Au spheres of 5 nm and 6.3 nm diameter. The Mie calculations were compared to experimental data in references [15, 31]. It was found in both cases that a good fit between calculated and measured data could not be obtained and further work is required in this area.

### 3.3. Final discussion

In summary, while previous works have modified a Drude term to account for the increase damping rate for small diameter Au spheres [15, 17-19, 31] this work developed a Drude-Lorentz dielectric function. The Lorentz terms in this phenomenological model are considered important to account for the intra-band transitions in the Au. The Drude-Lorentz function was obtained by a global optimization process using measurement data from Olmon et al [28]. The model was validated against published measured data for Au nanospheres and provides improved accuracy for diameters between than 10 nm and 30 nm. The model was seen to break down for spheres with diameter < 10 nm.

### 4. Conclusion

An improved Drude-Lorentz based material model for Au has been derived for sphere diameters between 10 nm and 30 nm. This model can be used to improve the accuracy of computational electromagnetic modelling of plasmonic components in the visual and near-infra red spectrum.

**References**

[1] H. Yu, Y. Peng, Y. Yang, and Z.-Y. Li, "Plasmon-enhanced light–matter interactions and applications," *npj Computational Materials,* vol. 5, no. 1, p. 45, 2019/04/11 2019, doi: 10.1038/s41524-019-0184-1.

[2] D. Darvill, A. Centeno, and F. Xie, "Plasmonic fluorescence enhancement by metal nanostructures: shaping the future of bionanotechnology," *Physical Chemistry Chemical Physics,* 10.1039/C3CP50415H vol. 15, no. 38, pp. 15709-15726, 2013, doi: 10.1039/C3CP50415H.

[3] S.-Y. Ding, E.-M. You, Z.-Q. Tian, and M. Moskovits, "Electromagnetic theories of surface-enhanced Raman spectroscopy," *Chemical Society Reviews,* 10.1039/C7CS00238F vol. 46, no. 13, pp. 4042-4076, 2017, doi: 10.1039/C7CS00238F.

[4] S. Lee, I. Choi, "Fabrication Strategies of 3D Plasmonic Structures for SERS," *BioChip Journal,* 2019/03/08 2019, doi: 10.1007/s13206-019-3105-y.

[5] C. E. Talley, J.B. Jackson, C. Oubre, N.K. Grady, C.W. Hollars, S. M. Lane, T.R. Huser, P. Nordlander, N.J. Halas, "Surface-Enhanced Raman Scattering from Individual Au Nanoparticles and Nanoparticle Dimer Substrates," *Nano Letters,* vol. 5, no. 8, pp. 1569-1574, 2005/08/01 2005, doi: 10.1021/nl050928v.

[6] M. Osawa, K.-I. Ataka, K. Yoshii, Y. Nishikawa, "Surface-Enhanced Infrared Spectroscopy: The Origin of the Absorption Enhancement and Band Selection Rule in the Infrared Spectra of Molecules Adsorbed on Fine Metal Particles," *Applied Spectroscopy,* vol. 47, no. 9, pp. 1497-1502, 1993/09/01 1993, doi: 10.1366/0003702934067478.

[7] J. Renger, R. Quidant, N. van Hulst, L. Novotny, "Surface-Enhanced Nonlinear Four-Wave Mixing," *Physical Review Letters,* vol. 104, no. 4, p. 046803, 01/29/ 2010, doi: 10.1103/PhysRevLett.104.046803.

[8] Z. Liu *et al.*, "Plasmonic nanoantenna arrays for the visible," *Metamaterials,* vol. 2, no. 1, pp. 45-51, 2008/05/01/ 2008, doi: https://doi.org/10.1016/j.metmat.2008.03.001.

[9] Z. Liang, J. Sun, Y. Jiang, L. Jiang, X. Chen, "Plasmonic Enhanced Optoelectronic Devices," *Plasmonics,* vol. 9, no. 4, pp. 859-866, 2014/08/01 2014, doi: 10.1007/s11468-014-9682-7.

[10] A. Boltasseva, H. A. Atwater, "Low-Loss Plasmonic Metamaterials," *Science,* vol. 331, no. 6015, p. 290, 2011, doi: 10.1126/science.1198258.

[11] K. R. Catchpole, A. Polman, "Plasmonic solar cells," *Opt. Express,* vol. 16, no. 26, pp. 21793-21800, 2008/12/22 2008, doi: 10.1364/OE.16.021793.

[12] T. Butburee, Z. Sun, A. Centeno, F. Xie, Z. Zhao, D.Wu, P. Peerakiatkhajohn, S. Thaweesak, H. Wang, L. Wang, "Improved $CO_2$ photocatalytic reduction using a novel 3-component heterojunction," *Nano Energy,* 2019/05/22/ 2019, doi: https://doi.org/10.1016/j.nanoen.2019.05.060.

[13] X. Cheng, S. Gu, A. Centeno, G. Dawson, "Plasmonic enhanced $Cu_2O$-Au-BFO photocathodes for solar hydrogen production," *Scientific Reports,* vol. 9, no. 1, p. 5140, 2019/03/26 2019, doi: 10.1038/s41598-019-41613-3.

[14] L. Wang, M. Hasanzadeh Kafshgari, M. Meunier, "Optical Properties and Applications of Plasmonic-Metal Nanoparticles," *Advanced Functional Materials,* vol. n/a, no. n/a, p. 2005400, doi: 10.1002/adfm.202005400.

[15] S. Berciaud, L. Cognet, P. Tamarat, B. Lounis, "Observation of Intrinsic Size Effects in the Optical Response of Individual Gold Nanoparticles," *Nano Letters,* vol. 5, no. 3, pp. 515-518, 2005/03/01 2005, doi: 10.1021/nl050062t.

[16] E. A. Coronado, G. C. Schatz, "Surface plasmon broadening for arbitrary shape nanoparticles: A geometrical probability approach," *The Journal of Chemical Physics,* vol. 119, no. 7, pp. 3926-3934, 2003, doi: 10.1063/1.1587686.

[17] A. Derkachova, K. Kolwas, I. Demchenko, "Dielectric Function for Gold in Plasmonics Applications: Size Dependence of Plasmon Resonance Frequencies and Damping Rates for Nanospheres," *Plasmonics,* vol. 11, no. 3, pp. 941-951, 2016/06/01 2016, doi: 10.1007/s11468-015-0128-7.

[18] A. Kheirandish, N. Sepehri Javan, H. Mohammadzadeh, "Modified Drude model for small gold nanoparticles surface plasmon resonance based on the role of classical confinement," *Scientific Reports,* vol. 10, no. 1, p. 6517, 2020/04/16 2020, doi: 10.1038/s41598-020-63066-9.

[19] P. N. Njoki, I. Im S. Lim, D. Mott, H-Y. Park, B. Khan, S. Mishra, R. Sujakumar, J. Luo, C-J. Zhong, "Size Correlation of Optical and Spectroscopic Properties for Gold Nanoparticles," *The Journal of Physical Chemistry C,* vol. 111, no. 40, pp. 14664-14669, 2007/10/01 2007, doi: 10.1021/jp074902z.

[20] H. Horvath, "Gustav Mie and the scattering and absorption of light by particles: Historic developments and basics," *Journal of Quantitative Spectroscopy and Radiative Transfer,* vol. 110, no. 11, pp. 787-799, 2009/07/01/ 2009, doi: https://doi.org/10.1016/j.jqsrt.2009.02.022.

[21] C. F. Bohren, D.R.Huffman, "Absorption and Scattering by an Arbitrary Particle," in *Absorption and Scattering of Light by Small Particles*, 1998, pp. 57-81.

[22] A. D. Rakić, A. B. Djurišić, J. M. Elazar, M. L. Majewski, "Optical properties of metallic films for vertical-cavity optoelectronic devices," *Appl. Opt.,* vol. 37, pp. 5271-5283, 1998, doi: https://doi.org/10.1364/AO.37.005271.

[23] M. N. Polyanski, Refractive index database [Online] Available: https://refractiveindex.info

[24] A. B. Djurisic, T. Fritz, and K. Leo, "Modelling the optical constants of organic thin films: impact of the choice of objective function," *Journal of Optics A: Pure and Applied Optics,* vol. 2, no. 5, pp. 458-464, 2000/07/31 2000, doi: 10.1088/1464-4258/2/5/318.

[25] B. K. Juluri. "Fitting Complex Metal Dielectric Functions with Differential Evolution Method." http://juluribk.com/?p=1597.

[26] B. J. Sumlin, W. R. Heinson, R. K. Chakrabarty, "Retrieving the aerosol complex refractive index using PyMieScatt: A Mie computational package with visualization capabilities," *Journal of Quantitative Spectroscopy and Radiative Transfer,* vol. 205, pp. 127-134, 2018/01/01/ 2018, doi: https://doi.org/10.1016/j.jqsrt.2017.10.012.

[27] P. B. Johnson, R. W. Christy, "Optical Constants of the Noble Metals," *Physical Review B,* vol. 6, no. 12, pp. 4370-4379, 12/15/ 1972, doi: 10.1103/PhysRevB.6.4370.

[28] R. L. Olmon, B. Slovick, T.W. Johnson, D. Shelton, S-H.Oh, G.D. Boreman, M.B. Raschke, "Optical dielectric function of gold," *Physical Review B,* vol. 86, no. 23, p. 235147, 12/28/ 2012, doi: 10.1103/PhysRevB.86.235147.

[29] K. M. McPeak, S.V. Jayanti, S.J.P. Kress, S. Meyer, S. Iotti, A. Rossinelli, D.J. Norris, "Plasmonic Films Can Easily Be Better: Rules and Recipes," *ACS Photonics,* vol. 2, no. 3, pp. 326-333, 2015/03/18 2015, doi: 10.1021/ph5004237.

[30] W. S. M. Werner, K. Glantschnig, C. Ambrosch-Draxl, "Optical Constants and Inelastic Electron-Scattering Data for 17 Elemental Metals," *Journal of Physical and Chemical Reference Data,* vol. 38, no. 4, pp. 1013-1092, 2009, doi: 10.1063/1.3243762.

[31] S. Karimi, A. Moshaii, S. Abbasian, M. Nikkhah, "Surface Plasmon Resonance in Small Gold Nanoparticles: Introducing a Size-Dependent Plasma Frequency for Nanoparticles in Quantum Regime," *Plasmonics,* vol. 14, no. 4, pp. 851-860, 2019/08/01 2019, doi: 10.1007/s11468-018-0866-4.

[32] A. F. Oskooi, D. Roundy, M. Ibanescu, P. Bermel, J. D. Joannopoulos, S. G. Johnson, "Meep: A flexible free-software package for electromagnetic simulations by the FDTD method," *Computer Physics Communications,* vol. 181, no. 3, pp. 687-702, 2010/03/01/ 2010, doi: https://doi.org/10.1016/j.cpc.2009.11.008.

[33] B. Dold, Mecke,R., "Optische Eigenschaften von Edelmetallen, Ubergangsmetallen und deren Legierungen im Infrarot (1. Teil)," (in German), *Optik* vol. 22, pp. 435-446, 1965.

[34] M.-L. Thèye, "Investigation of the Optical Properties of Au by Means of Thin Semitransparent Films," *Physical Review B,* vol. 2, no. 8, pp. 3060-3078, 10/15/ 1970, doi: 10.1103/PhysRevB.2.3060.